# The EU DataGrid Workload Management System: towards the second major release

G. Avellino, S. Beco, B. Cantalupo, A. Maraschini, F. Pacini, A. Terracina
*DATAMAT S.p.A*

S. Barale, A. Guarise, A. Werbrouck
*INFN, Sezione di Torino*

D. Colling
*Imperial College London*

F. Giacomini, E. Ronchieri
*INFN, CNAF*

A. Gianelle, R. Peluso, M. Sgaravatto
*INFN, Sezione di Padova*

D. Kouril , A. Krenek, L. Matyska, M. Mulac, J. Pospisil, M. Ruda, Z. Salvet, J. Sitera, M. Vocu
*CESNET*

M. Mezzadri, F. Prelz
*INFN, Sezione di Milano*

S. Monforte, M. Pappalardo,
*INFN, Sezione di Catania*

L. Salconi
*INFN, Sezione di Pisa*

In the first phase of the European DataGrid project, the 'workload management' package (WP1) implemented a working prototype, providing users with an environment allowing to define and submit jobs to the Grid, and able to find and use the "best" resources for these jobs. Application users have now been experiencing for about a year with this first release of the workload management system. The experiences acquired, the feedback received by the user and the need to plug new components implementing new functionalities, triggered an update of the existing architecture. A description of this revised and complemented workload management system is given.

## 1. INTRODUCTION

The European DataGrid project (EDG) [1] is a project funded by the European Union, with the aim to design and implement a Grid computing infrastructure, providing access to large sets of distributed computational and data resources, and suitable for the needs of widely distributed scientific communities. In the context of the EDG project, Work Package 1 [2] was mandated to build a suitable system for scheduling and resource management in a Grid environment.

During the first phase of the project, a Grid Workload Management System (WMS) was designed and implemented (also by integrating existing technologies), and deployed in the DataGrid testbed.

This first WMS, described in [3], has then been reviewed and complemented. In short the objectives to review the architecture of the WMS, discussed in this paper, were:

- to address the shortcomings that emerged in the first DataGrid testbed, in particular some scalability and reliability problems;
- to make it easy to plug-in new components implementing new functionalities;
- to favor the interoperability with other Grid frameworks.

The principles and the lessons learned when evaluating the first system on the DataGrid testbed, were applied when reviewing the architecture of the WMS: these are discussed in more detail in another CHEP 2003 paper [4].

In section 2 the new Workload Management System is presented. Section 3 describes the most significant improvements of this new WMS with respect to the first system, while section 4 discusses about some of the new introduced functionalities. In section 5 it is discussed about the foreseen future activities. Section 6 concludes the paper.

## 2. THE NEW WORKLOAD MANAGEMENT SYSTEM ARCHITECTURE

The new revised Workload Management System architecture is represented in Figure 1.

**MOAT007**

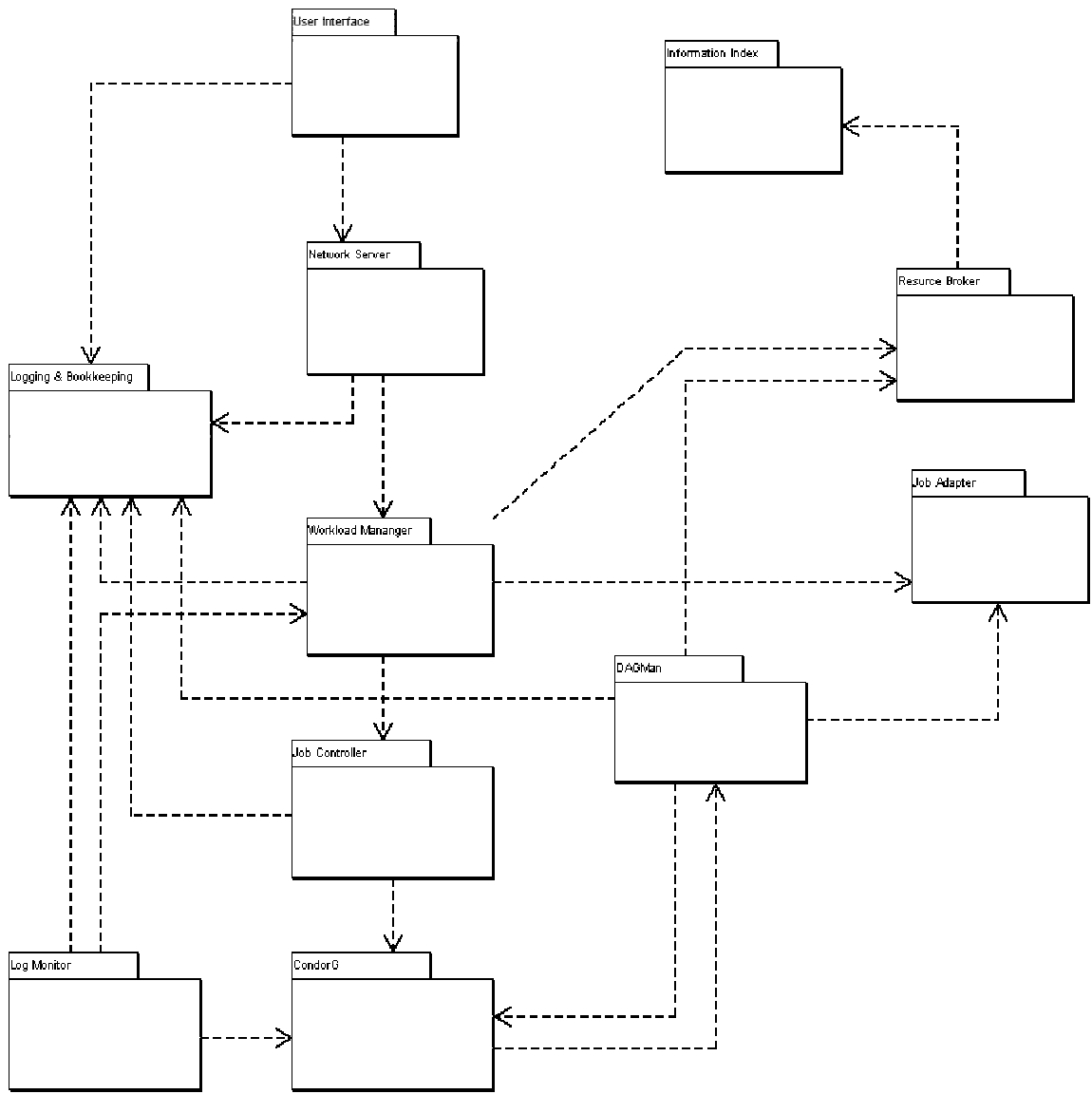

**Figure 1: The new WMS architecture**

As in the first release of the Workload Management System, the *User Interface* (*UI*) is the component that allows users to access the functionalities offered by the WMS. In particular, via the UI, users are allowed to submit jobs, which also includes the staging of some input files (the so-called input sandbox files) from the file system of the UI machine to the worker node where the execution will take place, to control them (cancel them, monitor their status), to retrieve the output files produced by the job (the so-called output sandbox files), etc.

Characteristics, requirements and preferences of jobs are specified via a Job Description Language (JDL), based on the Condor ClassAd language [5].

The *Network Server* is a generic network daemon, responsible for accepting incoming requests from the UI (e.g. job submission, job removal), which, if valid, are then passed to the Workload Manager. For this purpose the Network Server uses *Protocol*, to check if the incoming requests conform to the agreed protocol.

The *Workload Manager* is the core component of the Workload Management System. Given a valid request, it has to take the appropriate actions to satisfy it. To do so, it may need support from other components, which are specific to the different request types. All these components that offer support to the Workload Manager provide a class whose interface is inherited from a *Helper* class, which consists of a single method (*resolve()*). Essentially the Helper, given a JDL expression, returns a modified one, which represents the output of the required action. For example, if the request was to find a suitable resource for a job, the input JDL expression will be the one specified by the user at submission time, and the output will be the JDL expression augmented with the resource choice.

The *Resource Broker* (*RB*) or *MatchMaker* is one of these classes offering support to the Workload Manager. It is responsible to perform the matchmaking between the resource requirements (specified in the job JDL expression) and the status of the Grid. So, given a job submission request, the RB is responsible to find the resources that best match the request. For this purpose the RB has to interact with the Information Services, and also with the EDG Data Management services to resolve data requirements. The Resource Broker can be "decomposed" in three sub-modules:

- a sub-module responsible for performing the matchmaking, therefore returning all the resources suitable for that JDL expression;
- a sub-module responsible for performing the ranking of matches resources, therefore returning just the "best" resource suitable for that JDL expression;
- a sub-module implementing the chosen scheduling strategy, easily pluggable and replaceable with other ones implementing different scheduling strategies.

Within this architecture, the Resource Broker is therefore re-cast as a module, implementing the *Helper* interface, which can be "plugged" and used also in frameworks other than the EDG Workload Management System.

The *Job Adapter* is responsible for making the final "touches" to the JDL expression for a job, before it is passed to CondorG for the actual submission. So, besides preparing the CondorG submission file, this module is also responsible for creating the wrapper script: in fact the user job is wrapped within a script, which is responsible for creating the appropriate execution environment in the CE worker node (this includes the transfer of the input and of the output sandboxes).

*CondorG* [6] is the module responsible for performing the actual job management operations (job submission, job removal, etc.), issued on request of the Workload Manager. The CondorG framework is exploited for various reasons:

- the reliable two-phase commit protocol used by CondorG for job management operations, along with other provisions to increase the scalability of the GRAM protocol;

- the persistency: CondorG keeps a persistent (crash proof) queue of jobs;
- the logging system: CondorG logs all the relevant events (e.g. job started its execution, job execution completed, etc.) concerning the managed jobs: this is useful to increase the reliability of the whole system;
- the increased openness of the CondorG framework;
- the need for interoperability with the US Grid projects, of which CondorG is an important component.

The *Log Monitor* is responsible for "watching" the CondorG log file, intercepting interesting events concerning active jobs, that is events affecting the job state machine (e.g. job done, job cancelled, etc.), and therefore triggering appropriate actions.

For what concerns the *Logging and Bookkeeping* (*LB*) service, it stores logging and bookkeeping information concerning events generated by the various components of the WMS. Using this information, the LB service keeps a state machine view of each job. In the new WMS the LB is essentially the only job repository information. The dependencies between this component and the other modules of the Workload Management System (UI accessing the LB service to get status and logging information on jobs, and the various modules pushing events concerning jobs to the LB) are not represented in the figure, just for increased simplicity.

## 3. IMPROVEMENTS OF THE WORKLOAD MANAGEMENT SYSTEM

Various improvements were applied when designing the new WMS architecture, applying the lessons learnt while evaluating the first Workload Management System in the EDG testbed.

First of all the duplication of persistent information related to jobs (which was difficult to keep coherent, and which caused various problems) was avoided. As already mentioned, in the new WMS the LB service is essentially the only repository for job information. The drawback is that the reliability and scalability of this service is now much more important and critical than in the past. To address this issue, besides various improvements in the design and implementation, the new WMS has been designed to make possible relying on multiple LB servers per single WM, in order to distribute the load among multiple servers and therefore avoiding bottlenecks.

Another major improvement was the introduction of various techniques and capabilities to quickly recover from failures (e.g. process or system crashes). For example, the communication among the various components of the new WMS is now much more reliable, since it is done via persistent queues implemented in the file system.

In the new Workload Management System, moreover, monolithic long-lived processes were avoided. Instead,

some functionalities (e.g. the matchmaking) have been delegated to pluggable modules. This also helps reducing the exposure to memory leaks, coming not only from the EDG software, but also from the software linked with the WMS software.

Having the RB-Matchmaker as a pluggable module also increases the flexibility of the whole system, and the interoperability with other Grid frameworks. In fact it is now much more feasible to exploit the Resource Broker also "outside" the EDG Workload Management System. Moreover it is much more easier to implement and "plug" in the system the module implementing the chosen scheduling strategies, defined according to the one's own needs and requirements. Interoperability is also favored by the compliance of the new WMS to the Glue schema [7] [8], the common schema for information services agreed between European and US High Energy and Nuclear Physics Grid projects.

Other enhancements in design and implementation were applied to all the services, addressing the various shortcomings seen with the first release of the WMS. Improvements are also due to enhancements in the underlying software, such as the ones coming from the Globus [9] and the Condor [10] projects.

## 4. NEW FUNCTIONALITY

The new Workload Management System also implements some new functionalities, not available in the first release of the software.

Gangmatching is one of the new functionalities provided by the Resource Broker. It allows to take into account both computational and storage resources information in the matchmaking. So, for example, a user could specify that his jobs must be executed on a computational resource "close" to a storage system where there is "enough" free space available.

Job checkpointing is another of these new functionalities. Instead of addressing the classic checkpointing problem, that is saving somewhere all the information related to a process (process's data and stack segments, information about open files, pending signals, CPU state, etc.) as it is addressed in other projects (e.g. Condor [11]), the idea was providing users with a "trivial", or *logical* checkpointing service: through a proper API, a user can save, at any moment during the execution of a job, the state of this job. So users can insert in the code for their applications some specific function calls to save, from time to time, the state of their jobs. A checkpointable application must be able, of course, to restart itself from a previously saved state. In this "trivial" checkpointing service a state is defined by the user, and it is represented by a list of <var, value> pairs. They must allow to represent exactly what that job has done until that moment, and they must be chosen by the user in such a way that, relying on them, the job can restart later its processing from this intermediate state. This checkpointing framework is useful when a job is aborted because of an "external" problem (e.g. a machine crash), and in these cases the job is automatically rescheduled (possibly on a resource different than the one where the problem happened) and resubmitted. If a state for that job was saved in its previous execution, the job doesn't need to start from the beginning, but it can start from the "point" corresponding to the last saved state. Since it is not always so straightforward to "automatically" (by the Grid middleware) understand when a job ends in an "abnormal" way, it was also foreseen to allow the user to retrieve an intermediate state for a job (usually the last saved one), and explicitly resubmit the job, pointing out that it must start using this intermediate state.

For what concerns the architecture of the checkpointing framework, the functionality of persistently saving the state of a job, and of retrieving a previously saved state, is provided by the LB service.

The *DataGrid Accounting System* (DGAS) is another new functionality offered by the revised WMS. It is a closed economy based Grid accounting framework where users and resources are seen as entities capable of exchanging "virtual credits". For example, when a user submits a job to a Grid resource, the user pays to the resource a well-defined amount of credits in order to get the job executed. Generally a user receives the amount of credits needed to perform his computations by the management of the research group he belongs to. Research groups have their own Grid resources, and these resources earn credits by executing user jobs. These credits can then be redistributed among the users belonging to that group.

DGAS has two main purposes:

- Accounting for Grid Users and Resources
  It is possible to easily take tracks about resources used by the various users, and about the usage of the available Grid resources.
- Economic Brokering
  Help the Resource Broker in choosing the most suitable resource for a given job. In fact, once a valid price setting policy has been established, the model should lead to a state of nearly stable equilibrium able to satisfy the needs of both resource providers and consumers.

As first step, only the first functionality is provided by the new Workload Management System.

The new WMS also allows the execution of interactive jobs. This was done by integrating the Condor bypass software [12], making available a channel for the standard streams (stdin, stdout, stderr) from the worker node where the execution takes place to a remote machine, typically the User Interface machine, where the user can 'control' the job.

One of the most interesting new features of the WMS is the new extended querying capability of the Logging and Bookkeeping service. Users are allowed to define and mark jobs via user tags, and can then specify queries on these user tags and on the other "standard" fields. Just as example, a user could ask to get the status of all his jobs referring to production 'xyz' (user tag) and running on resource X or on resource Y.

In the new Workload Management System it is also possible to submit parallel (MPI) jobs, considering the MPICH implementation, a widely used, freely available, portable implementation of MPI.

Last but not least, it should be mentioned that in the new software release, it is possible to access the Workload Management functionalities not only via a python command line interface (as it was the case for the first release of the system), but also via C++ and Java API, and also via a Graphical User Interface (GUI).

## 5. FUTURE WORK

The new WMS also provides hooks for some other new functionalities that will be implemented and integrated later (actually the development of most of this software is already in good progress).

As already mentioned, the Economic Brokering, that is the integration of Grid Accounting with the Resource Broker (so that the most suitable resource for a given job is chosen according to the current price of resources and a pre-defined economic policy), is one of these new future capabilities of the WMS.

Another new functionality that will be provided is the support of inter-job dependencies, which can be defined by Directed Acyclic Graphs (DAGs), whose nodes are program executions (jobs), and whose arcs represent dependencies between them. Within the Workload Management System, a DAG will be managed by a meta-scheduler, called DAGMan (DAG Manager), whose main purpose is to navigate the graph, determine which nodes are free of dependencies, and follow the execution of the corresponding jobs. DAGMan is a product originally developed within the Condor project [10]. DAGMan can therefore be seen as an iterator through the nodes of a DAG, looking for free nodes (i.e. nodes without dependencies). The corresponding jobs can then be submitted for execution. Before doing this, it is of course necessary to choose the resource where to submit the job, and this will be done considering a lazy scheduling model, that is a job (node) is bound to a resource just before that job is ready to be submitted.

Job partitioning is another functionality that will be introduced in the Workload Management System framework. Job partitioning takes place when a job has to process a large set of "independent elements", as it often happens in many applications, such as most HENP applications. In these cases it may be worthwhile to "decompose" the job into smaller sub-jobs (each one responsible for processing just a sub-set of the original large set of elements), in order to reduce the overall time needed to process all these elements through "trivial" parallelisation, and to optimize the usage of all available Grid resources. The proposed approach is to address the job partitioning problem in the context of the logical job checkpointing framework described above: the processing of a job could be described as a set of independent steps/iterations, and this characteristic can be exploited, considering different, simultaneous, independent sub-jobs, each one taking care of a step or of a sub-set of steps, and which can be executed in parallel. The partial results (that are the results of the various sub-jobs) can be represented by job states (the final job states of the various sub-jobs), which can then be merged together by a job aggregator, which must start its execution when the various sub-jobs have terminated their execution.

Immediate or advance reservation of resources, which can be heterogeneous in type and implementation and independently controlled and administered, is another new functionality that will be supported, to allow the use of end-to-end quality of service (QoS) services in emerging network-based applications. The Workload Management System will provide a generic framework to support reservation of resources, based on concepts that have emerged and been widely discussed in the Global Grid Forum. In its implementation it is foreseen to address at least computing, network and storage resources, provided that adequate support exists from the local management systems.

## 6. CONCLUSIONS

The first Workload Management System, which was implemented in the first phase of the DataGrid project, and evaluated in the DataGrid testbed also in some quasi-production experiment activities, has been reviewed. The object was in particular to address some of the existing problems and shortcomings, and to support some new functionality. Moreover in the new WMS the hooks needed to implement some new functionalities, to be provided later, were implemented.

The preliminary results of the new Workload Management System, in terms of reliability, stability and performance are very encouraging. A more comprehensive evaluation will be possible when real test activities performed by real users on the large scale DataGrid testbed will be performed (at the time of writing, the new WMS is being integrated in this testbed).

## Acknowledgments

DataGrid is a project funded by the European Commission under contract IST-2000-25182.

We also acknowledge the national funding agencies participating to DataGrid for their support of this work.